# Drivers of carbon fluxes in Alpine tundra: a comparison of three empirical model approaches

Marta Magnani[a,b*], Ilaria Baneschi[c], Mariasilvia Giamberini[c], Pietro Mosca[a], Brunella Raco[c], Antonello Provenzale[c]

[a] Institute of Geosciences and Earth Resources, Via Valperga Caluso 35, 10125 Torino, Italy
[b] University of Turin & INFN, Via Pietro Giuria 1, 10125 Torino, Italy
[c] Institute of Geosciences and Earth Resources, Via Moruzzi 1, 56124 Pisa, Italy

* corresponding author: Marta Magnani, email: marta.magnani@edu.unito.it, address: Via Pietro Giuria 1, 10125 Torino, Italy

## *Abstract*

In high mountains, the effects of climate change are manifesting most rapidly. This is especially critical for the high-altitude carbon cycle, for which new feedbacks could be triggered. However, mountain carbon dynamics is only partially known. In particular, models of the processes driving carbon fluxes in high-altitude grasslands and Alpine tundra need to be improved. Here, we propose a comparison of three empirical approaches using systematic statistical analysis, to identify the environmental variables controlling $CO_2$ fluxes. The methods were applied to a complete dataset of simultaneous in situ measurements of the net $CO_2$ exchange, ecosystem respiration and basic environmental variables in three sampling sites in the same catchment. Large year-to-year variations in the gross primary production (GPP) and ecosystem respiration (ER) dependences on solar irradiance and temperature were observed,. We thus implemented a multi regression model in which additional variables were introduced as perturbations of the standard exponential and rectangular hyperbolic functions for ER and GPP, respectively. A comparison of this model with other common modelling strategies, showed the benefits of this approach, resulting in large explained variances (83% to 94%). The optimum ensemble of variables explaining the inter- and intra-annual flux variability included solar irradiance, soil moisture and day of the year for GPP, and air temperature, soil moisture, air pressure and day of the year for the ER, in agreement with other studies. The modelling approach discussed here provides a basis for selecting drivers of carbon fluxes and understanding their role in high-altitude Alpine ecosystems, also allowing for future short-range assessments of local trends.

**Highlights**

- Understanding the environmental variables that control $CO_2$ fluxes in high-altitude ecosystems is critical to assess the current and future changes in the mountain carbon cycle.
- A complete set of measurements of carbon fluxes and environmental variables in high-altitude grasslands and Alpine tundra is presented and analyzed.
- We propose a multi regression empirical model based on measured data that explains up to 90% of the carbon flux variability.
- Soil moisture and the day of the year (as a proxy of the vegetation seasonal cycle) are shown to act as relevant components in determining carbon fluxes.

# 1  Introduction

The carbon cycle plays a fundamental role in Earth's climate, as it controls the amount of greenhouse gases in the atmosphere and the carbon stocks in the soils, biosphere, oceans and atmosphere. A full budgeting of the carbon content in the Earth's reservoirs is still missing, owing also to the complexity of the driving processes. The carbon cycle spans a wide range of space and time scales, form the geological and slow dynamics associated with plate tectonics and surface rock weathering (e.g., Broecker, 2018; Orcutt et al., 2019) to the fast an much more intense fluxes generated by the action of living organisms. The fast cycle mainly develops in the near-surface terrestrial and marine environments, where ecosystems and the services they provide are "critical" to the survival of the whole biosphere.

The heterogeneous terrestrial environment at the Earth's surface, where "rock meets life", is tightly coupled with the water, carbon and energy cycles, providing essential ecosystem services such as water regulation and carbon sequestration. This part of the Earth System has recently been called the "Critical Zone" (CZ, NRC, 2001; Giardino and Houser, 2015). In the past decade, a dense network of CZ Observatories (CZO) was globally established (www.czen.org). Such observatories are often located in climate-change hotspots and along strategic gradients of altitude, latitude, climate, biome and habitat. In many of these sites, researchers pursue the integrated exploration of the complex processes linking geosphere, biosphere and climate (Rietkerk et al., 2011).

In this framework, mountain areas are gaining great attention owing to their rapid response to climate change and direct human impact. In 2013, the IPCC report (IPCC, 2013) documented a faster warming trend of the Alpine environment compared to the average north-hemisphere tendency. In mountains, the mean annual temperature increased of about $2^oC$ in the past century, with major intensification starting from the 1980s (Auer et al., 2007). This behaviour is an example of the elevation-dependent warming observed in many mountain areas of the world (Pepin et al., 2015). Changes in the hydrological cycle have also been recorded, including both intensification of winter and spring precipitation in some areas (Schmidli and Frei, 2005), and significant and progressive drying in others (Brunetti et al., 2006, 2009).

To contribute to the monitoring of climate change impacts in mountains, in 2017 an Alpine CZO (CZO@Nivolet) was established at the Nivolet Plain in the Gran Paradiso National Park (GPNP) in the western Italian Alps, see Figure 1A. Carbon dioxide fluxes and environmental variables were measured to estimate carbon storage and explore the drivers of carbon dioxide fluxes in high-altitude grasslands.

Grasslands are an important habitat as they cover nearly one-fifth of land surface in the world (Suttie et al., 2005) and are considered to be a relevant carbon sink (Janssens et al, 2005). However, a large variability is observed, depending on climate and location (e.g. Ciais et al, 2005). The primary drivers of $CO_2$ emission and uptake are usually taken to be the environmental temperature (Lloyd and Taylor, 1994) and the incident solar radiation (Ruimy et al., 1995), respectively. Precipitation was widely shown to strongly constrain those dependences (Fang et al., 2018). Aridity reduces productivity (Ciais et al., 2005) and inhibits emissions (Nagy et al.,2007), while bursts of emissions are observed after precipitation events that follows droughts (Flanagan et al., 2002; Wu et al, 2010; Xu and Baldocchi, 2004). A lagged dependence of the emissions on the productivity or the net uptake was also proposed, owing to the input of organic carbon enhancing bacterial activity (Balogh et al, 2011). Finally, mountain grasslands have been shown to act as stronger carbon sinks in comparison with their lower-elevation counterparts (Soussana et al., 2007; Gilmanov et al., 2007).

In past years, several multi regression models were adopted to include multiple predictors of carbon fluxes in grasslands. However, a single shared expression was hard to find. A common strategy relied on multiplicative models including, for instance, soil moisture as a constraint on temperature for ER (e.g. Reichstein et al, 2002; Zhao et al, 2019). Other studies used additive models to include further predictors, such as productivity (Balogh et al., 2011). In some cases, the regression parameters in the functional dependence of ER were suggested to depend on other variables (Reichstein et al, 2002; Xu and Baldocchi, 2004). Indirect drivers of geochemical origin, acting on the soil composition (Balogh et al., 2011; Emmerich, 2003), were also suggested to affect the processes controlling carbon fluxes.

In this work, our aim was twofold. First, we improved the representation of observed carbon dioxide fluxes by identifying an empirical model which explained a large part of the variance. To achieve this goal, we compared three methods of empirical modelling and we discussed their performances by means of a systematic statistical approach. We identified three sampling sites characterized by different parental geological materials within the same catchment in the CZO@Nivolet, and we compared the modelling results across the sites and in different years. The second aim was to improve knowledge on the behaviour of carbon drivers in high mountain environments, providing empirical information on carbon controlling processes to be implemented in future, more complete process-based models.

The rest of this paper proceeds as follows. In section 2 we introduce the materials and methods, with a characterization of the study area, a description of the measurement procedures and an explanation of the statistical methods used in data analysis and model building. Section 3 reports the results of the analysis and of the empirical modelling comparison, and section 4 is devoted to the discussion of the results. Section 5 gives conclusions and perspectives.

## 2 Materials and methods

### 2.1 Site description

The Critical Zone Observatory at Nivolet is located in the Gran Paradiso National Park (GPNP), see Figure 1. GPNP was established in 1922 for the protection of Alpine ibex (*Capra ibex*) and more generally of the Alpine ecosystems. It covers a total area of 720 km$^2$. The protected area includes mountainous terrain with large exposure of crystalline rocks, variably covered with glaciers and glacial deposits. It is characterized by typical Alpine woods at lower elevations and high-altitude grasslands and Alpine tundra above the tree line. The Nivolet Plain area is a glacial valley (Figure 1B), whose floor is located between 2600 m a.s.l. and 2300 m a.s.l.. Pleistocene unsorted till and glaciofluvial sediments with interbedded peat layers are preserved on the left flank of the valley. The bedrock consists of gneisses, dolostones and marbles belonging to the Gran Paradiso massif and calcschists with serpentinites and metabasites of the Piedmont-Ligurian zone (Figure 1B and 1C; e.g. Dal Piaz, 2010; Piana et al., 2017 and references therein).

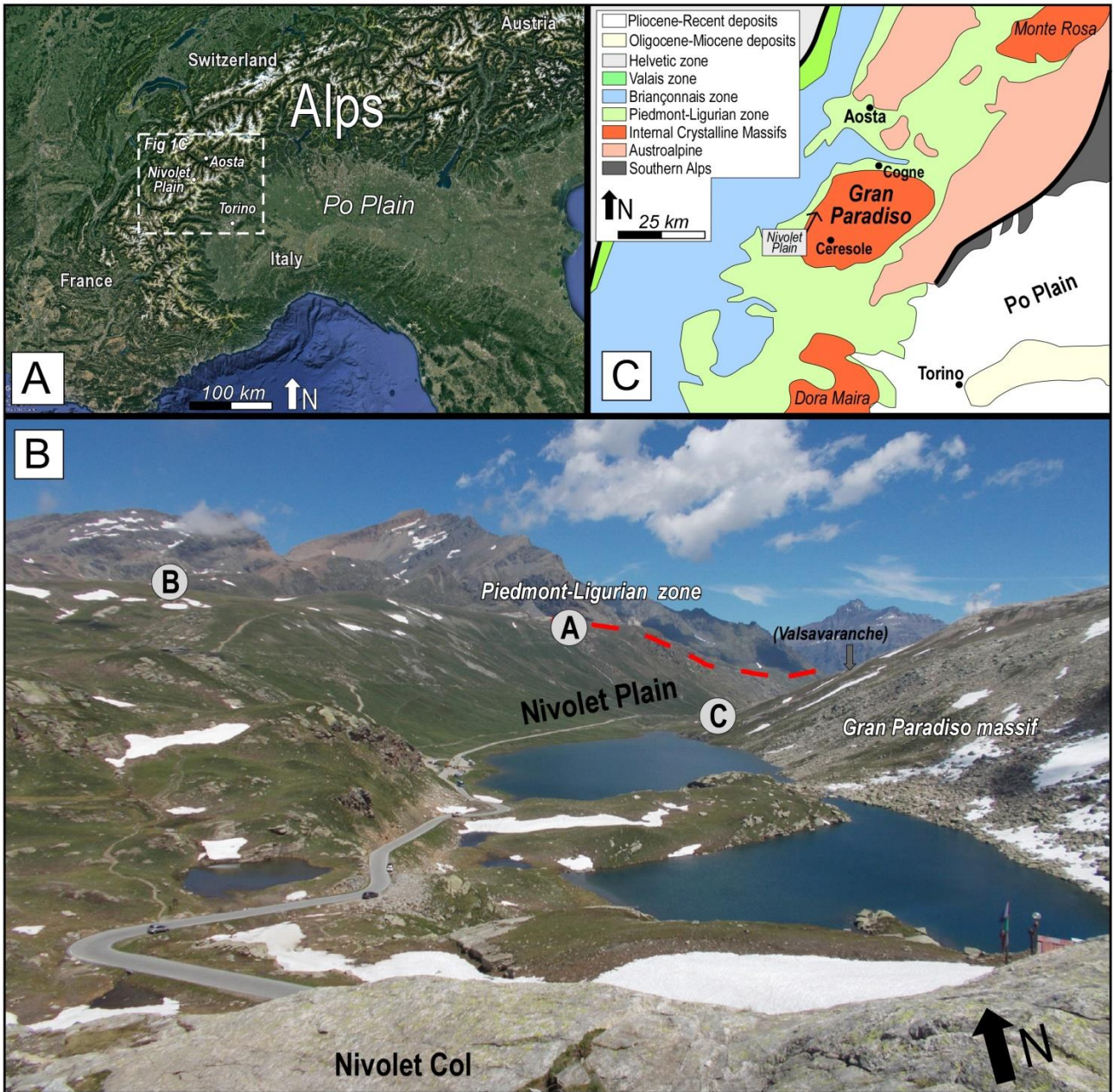

*Fig 1 – A) Location of the Nivolet Plain area in the western sector of the Alpine chain (image Landsat/Copernicus, @Google). B) View of the upper part of the Nivolet Plain from the Nivolet Col. The location of the three measurement sites (in the text reported as A-Carbonates, B-Glacial deposits and C-Gneiss) and the tectonic boundary between the Gran Paradiso massif and the overlying Piedmont-Ligurian zone are also shown. C) Simplified tectonic sketch map of the central part of the western Alps showing the location of the Nivolet Plain area.*

From November to early-mid June, the soil is covered with snow, with maximum snow depth exceeding 250 cm in the last three years, preventing *in-situ* measurements during wintertime. Daily records of precipitation, temperature and snow depth at the nearby Serrù weather station are available from 1962. The mean annual precipitation over the whole record is 1185 mm/year, with average precipitation during snow-free season (June-October) precipitation of about 492 mm/year. The mean daily temperature during June-October ranges from 3 to 12 °C (5th and 95th quantiles, respectively).

The Nivolet plain is covered with Alpine pasture, with typical high-altitude species such as *Carex, Trifolium, Silene* and *Geum* among the dominant species. The vegetation undergoes a fast phenological change during the summer months from mid-late June to October.

Three carbon measurement sites, each with an approximately square size of about 500 to 900 m$^2$, were identified taking into account the geological, geomorphological and environmental features of the area (Figure 1B). Two sites are located on the orographic left flank of the valley, facing South-East, and are characterized by soils on carbonate rocks (site A, at *2750 – 2760 m a.s.l.*) and on glacial deposits (Site B, at *2700 m a.s.l.*). The third site is located on soils developed on gneiss (site C, at about 2580-2600 *m a.s.l.*) along the orographic right flank of the valley. These three sites allowed for the comparison of $CO_2$ fluxes from different geological parental materials within the same watershed. Further details are reported in Table 1.

|  | Site A | | Site B | | Site C | |
|---|---|---|---|---|---|---|
| Coord. | Latitude N | Longitude E | Latitude N | Longitude E | Latitude N | Longitude E |
|  | 45.500212 | 7.152213 | 45.490167 | 7.139916 | 45.490256 | 7.149253 |
| Total area, m$^2$ | 900 | | 500 | | 500 | |
| Elevation (asl) | 2750 – 2760 | | 2740 – 2750 | | 2580 - 2600 | |
| Aspect | E-SE | | S-SW | | W-NW | |

*Table 1. Geomorphological features referred to the median point of the sites and the slope facing directions.*

## 2.2 Measurement procedure

In the three sites, the fluxes of carbon dioxide were measured each year from July to October using the enclosed-chamber method (Chiodini et al., 1998). The estimates of the fluxes were obtained by means of a transparent cylindrical polycarbonate accumulation chamber (height: 31.5 cm; area of the base: 363 cm$^2$) connected to a LI-COR LI-840 IR spectrophotometer mounted inside a hard case provided with a back-pack harness. To perform the measurements, we place the chamber over a stainless-steel collar which was inserted into the soil (to a depth of about 1 cm). After a first short transient period, a nearly linear increase/decrease of the $CO_2$ concentration was observed before the flux intensity starts to decrease. An acquisition time of about one minute was fixed after preliminary tests, excluding the initial transient period of about five to ten seconds. In the nearly linear regime, the concentration tendency (slope of concentration data vs. time) allowed to estimate the flux. A positive flux indicates $CO_2$ emission and a negative flux indicates $CO_2$ uptake.

In each of the three sites and for each measurement campaign, 20 to more than 30 randomly selected sample points inside the site area were measured in about two hours. As detailed in Section 3.1, estimating a stable average requires a minimum of 15 samples points. In windy or cloudy weather, on the other hand, the individual measurements could significantly fluctuate owing to the rapid passage of clouds. For this reason, we chose to increase the number of sample points; outliers could then be excluded in post processing to reduce the noise while preserving the constraint on the minimum number of points needed for stable statistics. A fixed number of sample points was then used in the analysis for each campaign.

For each sample point, the flux measurement was performed first with the transparent chamber and then with the chamber shaded by a dark cover. Between two consecutive measurements, ambient air was allowed to flow inside the chamber until the atmospheric $CO_2$ concentration was restored. The Net

Ecosystem Exchange (NEE) and the Ecosystem Respiration (ER) were then recorded in a short time interval. Assuming negligible variations of the environmental variables between the two consecutive measurements of NEE and ER, this procedure allows to obtain the Gross Primary Production (GPP) for each sample point, i.e. the net carbon fixation due to photosynthesis, as the difference of these two consecutive measurements (GPP=NEE-ER).

For each site, a total of 14 to 16 measurement campaigns were performed in summers 2017, 2018 and 2019. The fluxes were monitored every 10 days, regardless of weather conditions, in order to account for the natural meteorological variability. Given the location of the sites and the time needed to perform the measurements, each campaign over the whole of the three sites lasted for two consecutive days. The data quality was tested, in the expected range of flux values, by means of a laboratory calibration curve that allows to convert the temporal variation of the $CO_2$ concentration inside the chamber (ppmV/s) into the $CO_2$ flux ($\mu molCO_2/m^2/s$). Note that, given the altitude, the typical vegetation height never exceeded the collar level and no volumetric correction was needed for our sampling method and area (Morton and Heinemeyer, 2018).The raw dataset was quality controlled by statistical analysis of the frequency distribution and by the Rosner tests (Rosner, 1983) to identify potential outliers.

Together with fluxes sampling, basic meteo-climatic variables were recorded. Air humidity ($q$), air temperature ($Ta$), air pressure ($Pr$) and solar irradiance ($rs$) were measured at about 1.5 m above the ground at the centre of the plot with a thermoigrometer and a pyranometer LSI Lastem. Soil temperature ($Ts$) and soil volumetric water content ($VWC$), that were expected to have significant spatial heterogeneity, were measured at the base of the collar at a depth of about 5cm (see Appendix A for a detailed description of the measurement apparatus, instrument specifications, calibration and post-processing). Thus, we obtained a complete dataset (Giamberini et al., 2019, available on the Zenodo platform) for each campaign and for each site, composed of simultaneous carbon dioxide fluxes and environmental variables. Soil pits dug within the plot boundaries confirmed that the shallow vegetation was associated with a root zone thickness of about 20 cm. Moreover, previous work showed that the (potential) correlation between soil moisture and fluxes mainly interests the first horizon of 0-20cm (Fang et al., 2018). Further details are given in Appendix A.

## 2.3 Statistical analysis

The statistical analysis was performed in three steps:
1. Analysis of the mean values of the fluxes and the meteorological variables;
2. Univariate flux dependence on common, standard predictors;
3. Implementation of a comprehensive multi regression model.

Given the intrinsic small-scale variability of some variables (e.g. the local $VWC$), the individual sample points in each plot and for each measurement date were averaged and a standard deviation was assigned to each campaign, so that step 2 and 3 were performed on the averaged values. The same analysis was also applied to the raw (unaveraged) samples and, where present, discrepancies will be discussed in section 4.

The predictors were sampled during both the NEE and ER measurements, verifying that environmental variables did not change between the two consecutive NEE and ER measurements at each sample point. The only exception was solar irradiance, that could potentially fluctuate owing to the fast transition of localized clouds. For this reason, and since the photosynthetic process varies rapidly with incident light, the set of predictors considered for the regressions was that acquired during the transparent-chamber record. Note that the ER measurements do not directly depend on the instantaneous value of solar irradiance, since the chamber was shaded.

The significance of the regressions was obtained by a double-tailed test. To this purpose, a large number of surrogate pairs *(x$_i$, y$_i$)* was generated by random shuffling the samples, avoiding repetitions. This allowed to obtain a P-value, that was used to test the distribution of the regression coefficient against the null hypothesis of statistically uncorrelated variables (Jacobson et al., 2004). The tested null hypothesis assumed the observed value to be induced by random coincidence, with P representing the probability that the observed correlation was generated by random fluctuations. Hence, we require P<0.05 in order to accept the significance of the observed values. The same method was applied to assess whether the observed differences between the sites were significant. In this case, the differences of randomly shuffled pairs were obtained and their distribution was used to test the significance of the observed differences.

In the third step, multiple regression models were compared, and their representativeness was examined. This information was quantified by the explained variance, that is $(\sigma_y^2 - \sigma_{res}^2)/\sigma_y^2$, with $\sigma_{res}^2$ the residual variance. However, when dealing with multi regression models, the inclusion of more parameters may increase the variance explained by the model to the detriment of the model parsimony. The selection of the optimal model was then obtained using the Akaike Information Criterion (AIC, Akaike, 1974). This method allowed to identify the most efficient model. Indeed, the AIC measures the goodness of a statistical model based on a trade-off between the explained variance and parsimony (that is, with a penalty proportional to the number of free parameters, $k$). By this criterion, the empirical model having the lowest AIC should be preferred. For regressive models, the expression of the AIC simplifies to $AIC = N \log(\sigma_{res}^2) + 2k$, with $N$ the number of data points and $k$ the number of estimated parameters (Burnham and Anderson, 2002) Finally, the gaussianity of residuals was tested using the Lilliefors' test (Lilliefors, 1967) and potential heteroscedasticity was checked by a Bartlett's test (Bartlett, 1937). The statistical analysis was performed in MatlabR2018b, with regressions, Lilliefors' and Bartlett's test intrinsic functions, while a specific function was implemented for the estimate of P-value using the shuffling method.

## *2.4 Model derivation*

An exponential dependence of ER on the environmental temperature (Lloyd and Taylor, 1994) and a response of GPP to light intensity through a Michaelis-Menten function (Ruimy et al., 1995) are commonly assumed. In particular, the factor that is commonly used to explain the canopy production is the incident photosynthetic active radiation (PAR). Since the PAR is a fraction of the total measured solar irradiance, *rs*, with a constant ratio $PAR/rs$ (Ruimy et al., 1995; Flanagan et al., 2002), we replaced the usual dependence on PAR with one on the measured *rs*. Thus, the explicit set of functions reads

$$ER = a \, exp(b \, T) + \delta, \tag{1}$$

$$GPP = \frac{F_{max} \, \alpha \, rs}{F_{max} + \alpha \, rs} + \delta \tag{2}$$

Where *a* is a free parameter, corresponding to the respiration 0°C, *b* is related to the usual Q$_{10}$ factor (Lloyd and Taylor, 1994) as Q$_{10}$ = exp(10 b), and δ includes all the unmodeled dependences, as well as the stochastic contribution from Gaussian noise, and the temperature T is expressed in Celsius. The parameters in Equation (2) are the maximum photosynthetic flux for infinite light supply (*F$_{max}$*) and the apparent quantum yield ($\alpha$).

Using Equations (1) and (2), a larger residual variance was often observed owing to the influence of other factors. This required to enhance the model representativeness by including new explanatory variables in a multi regression model (MM). Among all possible choices, for both ER and GPP we considered three multi regression model types, that incorporate new variables. The simplest model (here called MM1) is an additive one (e.g. Fu et al, 2009; Frank et al, 2002). In MM1, all explanatory variables with their most suitable functional forms are summed, therefore when the function of one variable tends to vanish, the others are not affected and the flux may still be large, owing to the other dependences. Hence, the single factors act to increase the magnitude of the overall flux, and a decrease in all factors is needed in order to suppress of the fluxes. A general formulation of MM1 is given by

$$\begin{cases} ER = a\,exp(b\,T) + f_1(x_1) + f_2(x_2) + \ldots + \delta \\ GPP = \frac{F_{max}\,\alpha\,rs}{F_{max} + \alpha\,rs} + g_1(y_1) + g_2(y_2) + \ldots + \delta \end{cases}, \quad (3)$$

where $f_i, i = 1,2,\ldots$ and $g_i, i = 1,2,\ldots$ are functions of the remaining variables, besides the environmental temperature, T, and the solar irradiance, *rs*, in the first and second equation, respectively. Here, we marked the additional predictors and their functions by different letters in the two equations, in order to highlight that the two sets are independent of each other.

Alternatively, the different explanatory variables involved in the process may act as limiting factors for the process itself. This effect can be described by a multiplicative model, here called MM2, where the explanatory environmental variables (or their functions) multiply the specific functions (1) and (2), identified by univariate regressions. Hence, when one of the function lowers, it limits the overall effect of the other factors, whatever their magnitude is (e.g. Jia *et al.*, 2007; Zhao *et al.*, 2019). In this case the general formulation of MM2 reads

$$\begin{cases} ER = a\,exp(b\,T) \cdot f_1(x_1) \cdot f_2(x_2) \cdot \ldots + \delta \\ GPP = \frac{F_{max}\,\alpha\,rs}{F_{max} + \alpha\,rs} \cdot g_1(y_1) \cdot g_2(y_2) \cdot \ldots + \delta \end{cases}, \quad (4)$$

where the convention used for the symbols is the same as in Equation (3).

Finally, it has been shown that the parameters of the exponential function for ER may vary from site to site and from year to year, owing for example to variations in precipitation and soil moisture (Reichstein e*t al.*, 2002; Xu and Baldocchi, 2004). Therefore, the idea adopted here is to express the parameters in Equations (1) and (2) as functions of the other environmental variables, besides the known dependences on temperature for ER (Equation(1)) and on solar irradiance for GPP (Equation(2)). We call this class of models MM3. An effort in this direction was proposed by Reichstein e*t al.* (2002), that assumed the parameter in the exponent of Equation (1) to depend on the soil moisture. If the additional dependences are expressed as relatively small perturbations of constant parameters, Taylor expansions become possible. Along these lines, we perturbed the set of Equations (1) and (2). Preserving only the first order of the expansions, one obtains the following expression

$$ER = a(x_1, x_2)e^{b(x_1,x_2)T} \approx (a_0 + a_1 x_1 + a_2 x_2 + a_0 b_1 x_1 T + a_0 b_2 x_2 T)\,e^{b_0 T} + \delta. \quad (5)$$

where $x_1$ and $x_2$ represent any two generic variables among the remaining ones. Clearly, this formula can be easily generalized to an arbitrary number of variables. Here, factors of the kind $x_1 x_2$ or $x_1^2$ are discarded, since inconsistent with the perturbation hypothesis, they should be negligible, compared to the leading terms (in mathematical term, if the perturbation is of order $\varepsilon$, then such quadratic terms are of order $\varepsilon^2$). The same procedure was applied to Equation (2), that drove to

$$GPP = \frac{F_{max}(y_1, y_2)\alpha(y_1, y_2) rs}{F_{max}(y_1, y_2) + \alpha(y_1, y_2) rs} \approx$$
$$\approx GPP_0 (1 + A_1 y_1 + A_2 y_2) - GPP_0^2 \left(B_1 y_1 + B_2 y_2 + B_3 \frac{y_1}{rs} + B_4 \frac{y_2}{rs}\right) + \delta. \tag{6}$$

Here, $\delta$ again represents the residual unexplained processes, while $y_1$ and $y_2$ are two variables among the remaining ones, with the exception of *rs*. For shortness, from the Taylor expansion we redefined the various constants in expressions (6), as

$$GPP_0 = \frac{F_0 \alpha_0 rs}{F_0 + \alpha_0 rs},$$

$$A_1 = \frac{F_1 \alpha_0 + F_0 \alpha_1}{F_0 \alpha_0},$$

$$A_2 = \frac{F_2 \alpha_0 + F_0 \alpha_2}{F_0 \alpha_0}, \tag{7}$$

$$B_1 = \frac{\alpha_1}{F_0 \alpha_0},$$

$$B_2 = \frac{\alpha_2}{F_0 \alpha_0},$$

$$B_3 = \frac{F_1}{F_0 \alpha_0},$$

$$B_4 = \frac{F_2}{F_0 \alpha_0}.$$

The parameters reported in (7) were explicitly derived in Appendix B. The three types of multi regression models (MM1, MM2 and MM3) were then compared by means of their AIC and further details on the best performing version will be discussed in the following.

## 3 Results

### 3.1 Seasonal variations and mean values

Figure 2 shows the means over the individual sample points (dots), the standard deviations (coloured bars) and the 10th-90th quantiles (dark bars) of their for ecosystem respiration (ER), carbon uptake by vegetation (GPP) and net carbon exchange (NEE=GPP+ER), obtained for each measurement campaign and site. Here, negative values represent NEE dominated by the uptake by photosynthesis, while positive values represent NEE dominated by the $CO_2$ emissions. By definition, ER is strictly positive and GPP is strictly negative. All fluxes featured a clear seasonal trend. Starting from August, the ER and GPP experienced a prominent decrease in their magnitude. An asymmetric, bell-shaped temporal evolution over the period June to October was evident in some cases. As discussed by Janssens et al. (2001) and Law et al. (2002) for a variety of ecosystems, stronger ER usually matches larger GPP; this was also the case for the sites studied here. On the one hand, this feature may be induced by the enhanced canopy activity and the higher supply of fresh carbon to soil linked to vegetation physiology, for the sustenance of microbial communities (e.g. Balogh *et al*, 2011; Gavrichkova et al, 2018). On the other hand, the majority of the environmental variables show a clear seasonal trend, mainly induced by the natural cycle of the solar irradiance that can cause the simultaneous variation of the two fluxes, without any strict causal relation between GPP and ER. For instance, higher irradiance induces both larger GPP and higher temperatures, that in turn increase ER. In this case, the parallel variation of the fluxes would be induced by concurrent causes, instead of a direct effect of GPP on ER. Similar argument may also be drawn for other variables such as soil moisture and air pressure.

The variability of the fluxes within the site was obtained from the distribution of the values from the individual sample points (thus, incorporating both measurement uncertainties and small-scale spatial heterogeneity of the fluxes inside the site). In any case, the estimate of the mean and standard deviation for each site and campaign converged rapidly whenever more than about 15 sampling points were considered (see Figure C.1 in Appendix C). Larger variances of the fluxes were observed at the beginning of the growing season when the vegetation cover of the soil was more heterogeneous and the functional types varied stronger( the coloured bars in Figure 2 correspond to one standard deviation). At the end of summer, the soil was dry and only sporadic wilted bushes were present, a condition that explained the lower variances and the almost negligible NEE. The average values of the meteorological and environmental variables recorded during the measurement campaigns for each site are shown in Figure C.2 in Appendix C.

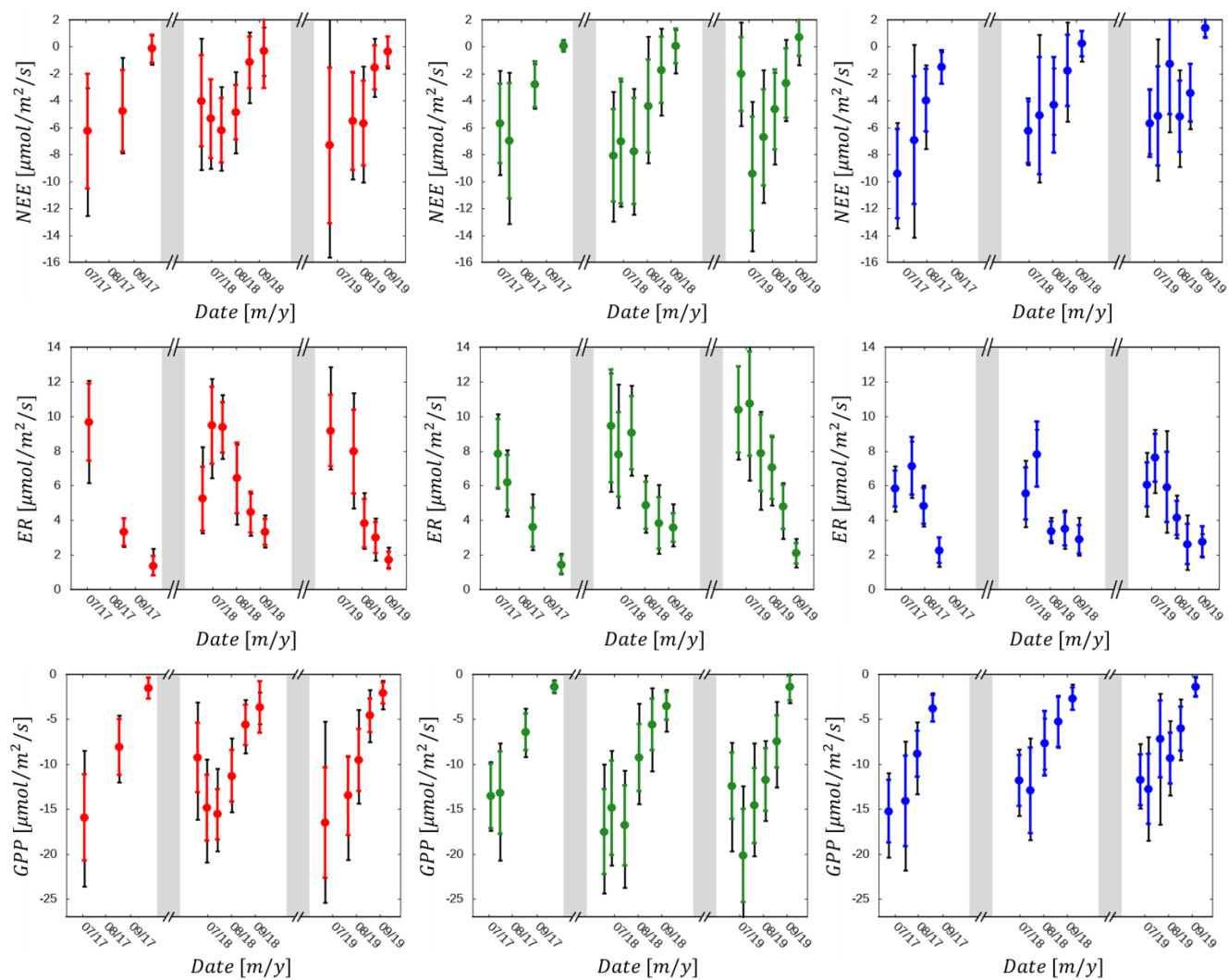

*Figure 2. Net Ecosystem Exchange (NEE, top), Ecosystem Respiration (ER, center) and Gross Primary Production (GPP, bottom) measured at site A (left, red), B (center, green) and C (right, blue) versus the measurement date. Dark arrows correspond to 10 and 90 quantiles and coloured bars indicate 1σ intervals.*

Overall, these Alpine meadows acted as a carbon sinks in all the three snow-free periods of measurements; plant fixation was about twice as large as ecosystem respiration.

The average values of the fluxes for the individual years and over the entire measurement period are reported in Table 2. Comparing the results for the three years, variations up to about 30% were evident between the average ER and GPP recorded in 2017, 2018 and 2019. One possibility is that such year-to-year variations were induced by changes in the environmental variables driving the fluxes. Therefore, the environmental variables recorded in the three sites were examined, looking for patterns that could possibly explain those of the fluxes. Moreover, a daily and a seasonal cycles of biosphere activity were expected, possibly inducing a discrepancy generated by a different hour and day of sampling. In order to account for such variations, the day of the year (*DOY,* 1-365d) and the hour of the day (*H,* 0-24h), were included as supplementary predictors in the analysis. Table 2 also reports the average values for the environmental variables in 2017, 2018 and 2019, and over the whole three-year period.

|  |  | $T_s$ [a] | $T_a$ [a] | VWC [b] | $R_s$ [c] | Q [b] | Pr [d] | H [e] | DOY [e] | NEE [f] | ER [f] | GPP [f] |
|---|---|---|---|---|---|---|---|---|---|---|---|---|
| 2017 | A | 20.6 | 19.0 | 14.9 | 840.4 | 30.0 | 739.0 | 12 | 247 | -4.3 | 5.5 | -9.8 |
|  | B | 20.7 | 19.0 | 21.7 | 758.1 | 35.0 | 740.2 | 13 | 239 | -4.5 | 5.5 | -10.0 |
|  | C | 19.9 | 19.3 | 24.4 | 837.4 | 37.4 | 752.1 | 13 | 223 | -6.3 | 5.8 | -12.1 |
| 2018 | A | 18.0 | 12.3 | 22.1 | 920.0 | 47.3 | 741.2 | 12 | 226 | -4.2 | 7.4 | -11.6 |
|  | B | 16.0 | 12.2 | 29.3 | 824.1 | 56.7 | 743.2 | 14 | 226 | -5.6 | 7.4 | -13.0 |
|  | C | 13.9 | 14.0 | 29.4 | 655.6 | 56.5 | 755.9 | 13 | 234 | -4.0 | 5.4 | -9.3 |
| 2019 | A | 11.9 | 10.0 | 17.7 | 635.1 | 63.0 | 739.9 | 12 | 233 | -4.7 | 6.0 | -10.7 |
|  | B | 12.8 | 10.9 | 29.2 | 769.9 | 61.7 | 741.4 | 13 | 228 | -5.0 | 8.5 | -13.5 |
|  | C | 11.3 | 11.7 | 23.1 | 599.6 | 65.9 | 754.6 | 14 | 229 | -3.9 | 5.7 | -9.6 |
| Average | A | 16.3 | 13.0 | 19.3 | 804.0 | 49.2 | 740.1 | 12 | 233 | -4.4 | 6.5 | -10.9 |
|  | B | 16.3 | 13.5 | 27.7 | 787.4 | 52.9 | 741.8 | 13 | 230 | -5.1 | 7.4 | -12.4 |
|  | C | 15.0 | 14.5 | 25.8 | 681.7 | 55.2 | 754.3 | 14 | 229 | -4.6 | 5.6 | -10.2 |

*Table 2. Mean values of the variables at the three sites (A, B and C) averaged over 2017, 2018, 2019 and on all three years ("Average").*

Most of the meteorological variables showed a clear seasonal trend. As expected, the mean daily solar irradiance and the soil temperature decreased starting from August, while a larger variability affected the air temperature (see Figure C.2 in Appendix C). The pressure recorded in the three sites mirrored the altitude distribution, i.e. site A and C corresponded to the lowest and highest mean pressure, respectively, with fluctuations around the mean values. The melting snow at the beginning of the summer season and the stronger rainfall of late spring increased the reservoir of soil water. Thereafter, the intense ecosystem activity determined a large water demand that was not balanced by the sporadic summer rainfall, resulting in a progressive decrease of soil humidity over the summer months.

The year-by-year means showed significant differences between years, as reported in Table C.1 in Appendix C. The summer of 2017 had the highest temperature and the lowest soil moisture. This identified the first year of sampling as the warmest and driest of the study period.

Comparing the three sites and averaging over the three years, site C showed the lowest ER and GPP values, (5.62±0.13 $\mu mol/s/m^2$ and -10.18±0.33 $\mu mol/s/m^2$, respectively), while the highest values were observed at site B (7.36±0.20 $\mu mol/s/m^2$ and -12.42±0.41 $\mu mol/s/m^2$, respectively). The values of ER displayed significant differences between the fluxes measured at the different sites over the whole period, while the differences for NEE were non-significant (Table C.2 in Appendix C). Higher $Ta$ and lower $VWC$ were measured on sites C and A, respectively, but the significance of the differences varied from year to year. These differences were probably induced by the different micrometeorological settings of the sites. Site A is located on a steep slope at the margin of the basin, where winds and drainage lower the local soil humidity . On the other hand, C is close to the valley floor, where a weaker air circulation leads to higher air temperatures. The similarity of air temperatures at A and B in 2017 and 2018 was striking and it combined with a similar behaviour of ER. However, no strict correlation between them can be stated considering the three years. In general, significant differences were observed in $VWC$ and $Pr$ in the three years. Differences in the incident solar radiation with site B were always significant, while a significant difference between A and C was present only in 2018.

In the following we investigate the dependence of the fluxes on the environmental variables, considering the mean values at each site and for each measurement campaign.

## 3.2 Standard nonlinear dependences

The soil (or air) temperature and the photosynthetic active radiation are well known drivers of ER and GPP, respectively. Therefore, as a first step, we analysed the dependences on these two variables, as shown in Figure 3 and 4. We used the air temperature as the explanatory variable for ER; similar results were obtained using the soil temperature.

The scatter plot of the ER vs $Ta$ is shown in Figure 3. A strong variation of the data points is evident. The data measured in 2017 are clearly separated from the 2018 and 2019 values . A single regression curve was hard to find and large residual variances were obtained. This is evident in Table 3, where the explained variance obtained considering the whole period of three years was, in general, lower than those of the separate years for all sites. Since 2017 was the warmest and driest year, these results suggest an effect of other meteorological variables on the expected dependence of ER.

| Site | Year | a [$\mu mol/m^2/s$] | b [$°C^{-1}$] | AIC | $\sigma^2_{expl}$ |
|---|---|---|---|---|---|
| A | 2017 | 0.02 | 0.29 | 3.38 | 0.19 |
|   | 2018 | 0.91 | 0.16 | 1.71 | 0.37 |
|   | 2019 | **0.67** | **0.02** | **-5.99** | **0.87** |
|   | All | 3.90 | 0.03 | 2.77 | 0.10 |
| B | 2017 | **0.006** | **0.35** | **-4.14** | **0.87** |
|   | 2018 | 1.65 | 0.12 | 2.34 | 0.25 |
|   | 2019 | **2.14** | **0.11** | **-4.45** | **0.76** |
|   | All | 5.90 | 0.016 | 3.63 | 0.03 |
| C | 2017 | **0.42** | **0.13** | **-15.60** | **0.99** |
|   | 2018 | 2.113 | 0.065 | 3.37 | 0.12 |
|   | 2019 | **1.81** | **0.09** | **-9.00** | **0.89** |
|   | All | **2.74** | **0.048** | **-2.16** | **0.34** |

*Table 3. Regression parameter of Equation (1) for Ecosystem Respiration, ER, AIC and explained variance over different years and the whole campaign (All). Significance of the regressions was evaluated by the shuffling method described in Section 2.4 and values having P<0.05 are highlighted in bold.*

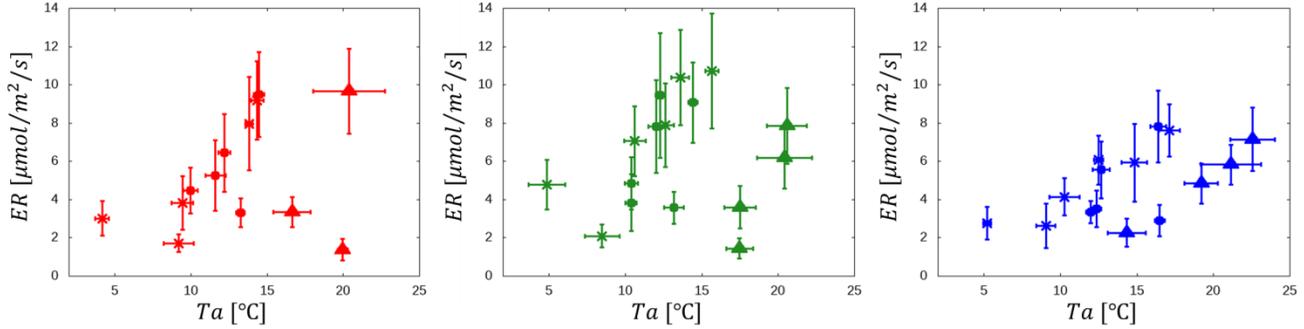

*Figure 3. Scatterplot of the ER and air temperature, for site A(left), B(center) and C(right). The three years of measurements are represented by different symbols, namely triangles (2017), circles (2018) and stars (2019).*

On the other hand, the GPP vs *rs* relationship did not display any strong year-to-year variability between years (Figure 4). A strong dispersion was however observed in campaigns performed from the end of September to the beginning of October, that correspond to the lowest GPP values. During those months, carbon uptake was almost uncorrelated with *rs*, presumably owing to the presence of scarce and wilting vegetation. At the end of the vegetative season, most of the measured plots were characterized by the prevalence of bare soil within the sampled surfaces, i.e. within the metallic ring. In such conditions, the GPP dependence on *rs* may be hampered by other factors. The corresponding flux values (marked in black in Figure 4) were thus excluded from the univariate regression (2).

| Site | Year | $F_{max}$ [$\mu mol/m^2/s$] | $\alpha$ [$\mu mol/W/s$] | AIC | $\sigma^2_{expl}$ |
|---|---|---|---|---|---|
| A | 2017 | 3.33 | -0.003 | -3.71 | 0.18 |
|   | 2018 | -107.74 | -0.019 | -7.25 | 0.30 |
|   | 2019 | -25.17 | -0.050 | -4.67 | 0.08 |
|   | All | -36.19 | -0.031 | -17.03 | 0.18 |
| B | 2017 | 5.28 | -0.0045 | -12.30 | 0.23 |
|   | 2018 | **-24.96** | **-0.010** | **-18.63** | **0.80** |
|   | 2019 | -7.28 | 0.014 | -9.17 | 0.28 |
|   | All | -35.16 | -0.036 | -14.34 | 0.23 |
| C | 2017 | 8.98 | -0.0064 | -5.45 | 0.33 |
|   | 2018 | -51.08 | -0.024 | -7.82 | 0.81 |
|   | 2019 | -25.62 | -0.037 | -8.13 | 0.60 |
|   | All | **-33.97** | **-0.027** | **-16.77** | **0.53** |

*Table 4. Regression parameter of Equation (2) for Gross Primary Production, GPP, AIC and explained variance over different years and the whole campaign (All). Significant regressions(P<0.05) are highlighted in bold. Measurements performed later than 15 September (black symbols in Figure 4) were excluded to achieve significant regressions.*

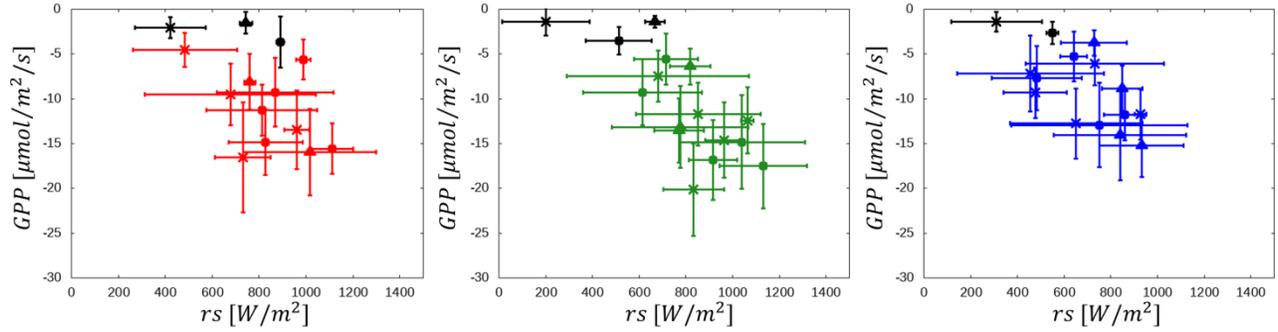

*Figure 4. Scatterplot of GPP versus solar irradiance, for site A(left), B(center) and C(right). The three years of measurements are represented by different symbols, namely triangles (2017), circles (2018) and stars (2019). Black point are values sampled after 15 September of each year.*

The above observations suggest that some additional processes should be included to properly represent the variability of the $CO_2$ fluxes. This drove us to include other variables in the model, moving to a multi regression approach.

### 3.3 Multi regression models

To model the year-to-year deviations from the standard functional forms, we included additional environmental variables in a multi regression approach, until a minimum AIC was achieved. A seasonal trend, present in many environmental variables, induced a correlation with the *DOY,* as it is shown by the correlation matrix of the predictors (see Table C.3 in Appendix C). This might bring to the use of the *DOY* in place of drivers. To avoid such issue, the *DOY* was included as the last predictor in the model. That is, the effect of the explicit *DOY* inclusion was considered only after the regression was already estimated with all the other measured predictors and the value of *H*.

In the multi regression test, we considered the whole data set for all three years; different combinations of the predictors were explored for each model (see Table C.4 for the AIC values and explained variances, in Appendix C). In MM1, the largest explained variances were obtained using *[Ta, VWC, DOY]* and *[rs, VWC]* in the regressions of ER and GPP, respectively. A linear dependence was supposed for VWC and a Gaussian function was used to reproduce the evolution with the DOY. However, the complexity of the model made the AIC to rise steeply with the introduction of the additional variables, resulting in the selection of the univariate fit. This means that, for our dataset, the model MM1 was rejected.

In MM2, the additional variable identified by the analysis was VWC, for both ER and GPP. The addition of other variables increases the complexity of the model without a significant benefit in the explained variance. Thus, in the final formulation of MM2, as selected by the AIC, the functions (1) and (2) were multiplied by the soil moisture. In both MM1 and MM2 either a Gaussian or a linear function of *DOY* were considered, with similar results. This is mainly due to the specific time window that we sampled, where the bell-shaped dependence observed in other works (e.g. Xu and Baldocchi, 2004) was uneven.

The most efficient model with the largest explanatory power turned out to be MM3. In this case, the lowest AICs were obtained with the sets *[Ta, VWC, Pr, DOY]* and *[rs, VWC, DOY]* for ER and GPP, respectively. The other possible predictors, including for instance the time of day when the measurements were taken, did not emerge as significant predictors. Despite the quite complicated formulation of Equation (6), analysis on the individual parameters indicated that all *B*-parameters were not significant in the regressions for any site. Similarly, all *b*-parameters in Equation (5) did not contribute to the improvement of the model. These findings allowed to simplify formulas (5, 6) as

$$\begin{cases} ER = (a_0 + a_1 VWC + a_2\,Pr + a_3\,DOY\,)\,e^{b_0 T_a} + \delta & (8) \\ GPP = \dfrac{F_0 \alpha_0 rs}{F_0 + \alpha_0 rs}(A_0 + A_1 VWC + A_2\,DOY) + \delta & (9) \end{cases}$$

Both equations retained the zero and first order terms in the Taylor expansions. The lowest (zero) order corresponds to the usual functional dependence while the additional predictors relapse at the first order. Moreover, here we added a free parameter, $A_0$, that was set to *1* in Equation (6). Such parameter accounts for the possible prevalence of the mixed dependences, such as $A_1 \dfrac{F_0 \alpha_0 rs}{F_0+\alpha_0 rs} VWC$ and $a_1 VWC\,e^{b_0 T_a}$, over the single dependences, $A_0 \dfrac{F_0 \alpha_0 rs}{F_0+\alpha_0 rs}$ and $a_0 e^{b_0 T_a}$ corresponding to the generalized standard functions of GPP and ER, respectively. Both $A_0$ and $a_0$ can vanish or become subdominant in the summation.

The parameters obtained for the three sites are reported in Tables 5 and 6. All the residuals had a Gaussian statistics at 95% significance level, according to the Lilliefors' test (Lilliefors, 1967). Moreover, the heteroscedasticity of the residuals in the range of variation of the data and across the three years was controlled at 95% significance level with Bartlett's test (Bartlett, 1937).

|   | $a_0$ [μmol/m²/s] | $a_1$ [μmol/m²/s] | $a_2$ [μmol/ m²/s/hPa] | $a_3$ [μmol/m²/s/d] | $b_0$ [°C$^{-1}$] | AIC | $\sigma^2_{expl}$ |
|---|---|---|---|---|---|---|---|
| A | -51.6 | -0.053 | 0.09 | -0.050 | 0.06 | -10.59 | 0.85 |
| B | -91.58 | 0.068 | 0.16 | -0.096 | 0.06 | -25.86 | 0.94 |
| C | -77.93 | 0.005 | 0.12 | -0.038 | 0.02 | -13.41 | 0.83 |
| All | -50.72 | 0.042 | 0.04 | -0.070 | 0.006 | -41.70 | 0.73 |

Table 5. Best fit parameters of Equation (8), AIC and explained variance ($\sigma^2_{expl}$) for the three sites. The last line ("All") was obtained by fitting all three sites together. All results refer to the full three-year data set.

|   | $F_0$ [μmol/m²/s] | $\alpha_0$ [$\mu mol$/W/s] | $A_0$ | $A_1$ | $A_2$ [1/d] | AIC | $\sigma^2_{expl}$ |
|---|---|---|---|---|---|---|---|
| A | -87.85 | -0.022 | 4.09 | -0.020 | -0.013 | -13.92 | 0.84 |
| B | -24.44 | -0.101 | 2.25 | 0.006 | -0.008 | -19.45 | 0.88 |
| C | -20.18 | -0.069 | 2.97 | 0.007 | -0.011 | -20.31 | 0.89 |
| All | -18.29 | -0.039 | 3.82 | 0.004 | -0.013 | -61.88 | 0.83 |

Table 6. Best fit parameters of Equation (9), AIC and explained variance ($\sigma^2_{expl}$) for the three sites. The last line ("All") was obtained by fitting all three sites together. All results refer to the full three-year data set.

Finally, Figure 5 shows the predicted (by Equations (8) and (9)) and measured flux values, for ER and GPP and for the three sites. A very good agreement between the model results and the measurements was obtained. Note, also, that some of the parameter values displayed a significant (P<0.05) difference between the three sites, as reported in Table C.5 in Appendix C. Significant differences are observed in the coefficients of the *VWC* dependence, in particular between sites A and B for ER and between site A and both B and C for GPP (with very significant differences).

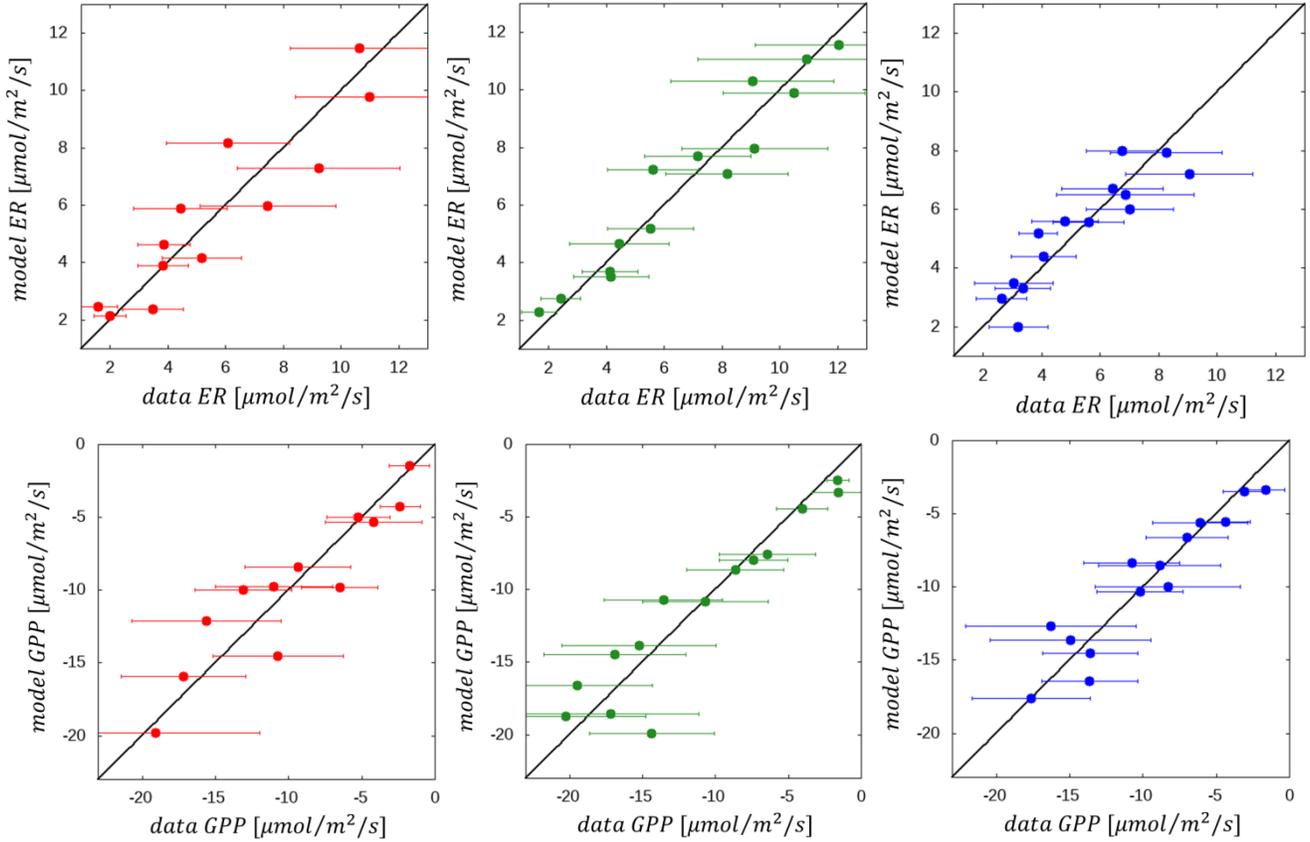

*Figure 5. Modelled versus measured fluxes. The upper and lower panels refer to ER and GPP, respectively. Site A, B and C are on the left, center and right. Error bars indicate the 1σ interval for the distributions of individual measured values.*

## *3.4 Sensitivity Analysis*

   The analysis discussed above was performed with *in-sample* regressions, i.e. all the available data were used to establish the model. However, the same procedure can be applied to just a portion of the dataset and the data that were not included in the calibration can be used to validate the predictive ability of the model. This is an *out-of-sample* prediction. Here, the strong difference in ER between different years suggested to split the dataset according to the year of sampling, using two years to calibrate the model parameters and the remaining year to compare modelled fluxes with measured ones (see Table C.6 and C.7 for ER and GPP, respectively, in Appendix C). Larger Root Mean Square Error, RMSE, were obtained in the projection of ER in 2017 using models calibrated on 2018 and 2019 data. Conversely, when 2017 and either 2018 or 2019 data were used to establish the model for ER, a good prediction of the remaining year was obtained. Clearly, the parameters obtained from 2018 and 2019, that had a similar behaviour in the ecosystem respiration and the values of the drivers, were not able to properly predict the large deviations of 2017. On the other hand, no significant difference associated with the choice of the calibration years was observed for GPP.
   Finally, as a proof of concept, we applied the model that was calibrated over the three years to obtain preliminary projections under simple "what-if" scenarios of climate change. We explored the effect of an increase of the mean temperature of 0.5 or 1 °C and of a possible positive or negative 10% variation in soil moisture. To this purpose, a constant term (offset) was added to the temperature and soil moisture drivers in Equations (8) and (9), increasing or decreasing their means while keeping the same variance of temporal fluctuations. From these, the percentage variations of the fluxes with respect to the average values measured in 2017-2019 were obtained. As expected, an increase in the ER was observed with rising temperature (see Tables 7 and 8). On the other hand, since soil moisture acts as a limiting driver for both ER and GPP in sites B and C, a decrease in the soil moisture under a warmer climate might mitigate the ER increase. The net effect depends on the relative magnitude of the ER and GPP variations, i.e. on their ratio. Indeed, while a drier soil lowers the carbon release via respiration, it also limits the plant $CO_2$ uptake by primary production, making the assessment of their balance non-trivial. Note that site A showed opposite trends owing to the negative sign of the parameters $a_1$ and $A_1$.

|  |  | A | | | B | | | C | | |
|---|---|---|---|---|---|---|---|---|---|---|
|  |  | $\Delta T$ | | | $\Delta T$ | | | $\Delta T$ | | |
|  |  | +0°C | +0.5°C | +1°C | +0°C | +0.5°C | +1°C | +0°C | +0.5°C | +1°C |
| $\Delta VWC$ | -10% | +0.09 | +0.12 | +0.14 | -0.008 | -0.014 | -0.019 | -0.02 | +0.025 | +0.037 |
|  | +0% | 0 | +0.07 | +0.09 | 0 | +0.017 | +0.023 | 0 | +0.028 | +0.040 |
|  | +10% | -0.003 | +0.02 | +0.04 | +0.06 | +0.06 | +0.05 | +0.14 | +0.032 | +0.043 |

*Table 7. Mean percentage variation (%) of the ER resulting from an increase in the mean temperature of 0,0.5,1 °C and from an increase/decrease in the mean soil moisture of +/-10%.*

|  |  | A | B | C |
|---|---|---|---|---|
| $\Delta VWC$ | -10% | +0.14 | -0.08 | -0.05 |
|  | +10% | -0.02 | +0.17 | +0.13 |

*Table 8. Mean percentage variation (%) of the GPP resulting from an increase/decrease in the mean soil moisture of +/-10%.*

# 4 Discussion

We applied a systematic statistical approach to establish a *quantitative* regression model that explained up to almost 90% of the $CO_2$ flux variability. In the regressions, similar results were obtained by using either soil or air temperature. Soil temperature suffered from larger fluctuations, resulting in lower, but compatible, explained variances. For such reason, we used *Ta* that appeared to be more stable and representative of the prevailing environmental conditions. For completeness, the whole analysis was repeated also using the whole set of measurements obtained from the individual sampling points, rather than the site averages, and the results described above were confirmed.

The statistical analysis revealed a strong dispersion of the measurements in the different years. A single exponential temperature dependence and a rectangular hyperbolic light response could not be satisfactorily fit (with a good representativeness and for the whole 3-yr period) to ER and GPP, respectively. In particular, within the same range of temperatures, ER was weaker in 2017 than in the two following years at all sites. Correspondingly, the average values of the environmental variables in the different years showed that 2017 had the warmest and driest summer. This suggested that the usual dependences were entangled by other processes, i.e. other environmental variables had to be included in a coherent model able to capture the inter-annual variability.

The complete set of environmental variables allowed the implementation of multi regression models for ER and GPP. We tested different models, considering functional forms that could be reciprocally independent (additive models, MM1), mutually limiting (multiplicative models, MM2) or that include the possible dependence of the parameters of the standard functions on other variables (parametric models, MM3). The results of the analysis indicated that MM3 provided the best description.

In all models, VWC appeared to be the most important perturbation to the fluxes, both for ER and GPP. This appears as a common behaviour in mountain areas, affecting ER in particular, as Xu and Baldocchi (2004), Fang et al. (2018) and Zhao et al. (2019), among others, documented in grasslands, steppe and meadows around the world. Our results indicate the need for explicitly including VWC in GPP models. Water availability can act as a limiting factor also for GPP, at least on high-altitude alpine environments, where soil moisture can become scarce during at least part of the vegetative season.

In the model for ER, pressure emerged as one of the environmental drivers affecting the dependence on temperature. The perturbation induced by pressure may be due to the leaching effect of $CO_2$ from soil to the atmosphere induced by the atmospheric pressure, similarly to the flux enhancing effect associated with wind (e.g. Emmerich, 2003; Spittlehouse and Ripley, 1977). However, a deeper exploration of the role of air pressure is beyond the scope of this work.

Since the values of the parameters $a_0$ and $A_0$ were non-null, a basal value of the fluxes was present, that depended only on the temperature (for ER) and incident solar radiation (for GPP). The perturbations induced by VWC, DOY and pressure (for ER) acted over these values. This maintains a hierarchy in the relevance of the predictors, ranking temperature and light as the most important predictors.

The parameter $b_0$ was used to obtain the usual $Q_{10}$ factor (Lloyd and Taylor, 1994), allowing a comparison of our results with other published values. Sites A and B showed the same value of $b_0$, as reported in Table 5. A significant (P<0.05) lower value of this parameter was found for site C. This mirrored the air temperature difference obtained between sites A, B and site C. The corresponding $Q_{10}$ values were 1.8 and 1.2 for A (or B) and C respectively, well within the range of published values for grasslands and steppe (e.g. Reichstein et al., 2002).

In MM3, a dependence on the *DOY* was still present for both ER and GPP after taking into account the measured environmental variables. According to the AIC, none of the other predictors was able to replace the *DOY*. The value of *DOY* thus emerged as a proxy for the effect of unmeasured environmental factors inducing a seasonal dependence in the fluxes, and/or for the physiological state of vegetation. To further explore this point, model MM3 was tested on the linearly detrended version of the variables, excluding the *DOY*. The same forms of Equations (8) and (9) were again found, identifying the same optimal set of explanatory variables (clearly without the *DOY*), although with a lower total explained variance (of the order of 40%). This suggested that two main components determine the fluxes: (1) a seasonal trend and (2) superposed sub-seasonal and inter-annual fluctuations which can be explained by predictors such as temperature, light, soil moisture and pressure. Clearly, the "snapshot" character of our sampling procedure did not allow for fully representing short-term, sub-seasonal fluctuations. These could better be analysed with a continuous measurement strategy, such as that provided by eddy covariance, but with the drawback of focusing on a pre-selected fixed spatial location.

A regression model was also obtained by merging the measurements at the three sites (last row of Tables 5 and 6). In this case, the explained variance of ER decreased, while the explained variance of GPP remained comparable with those obtained therefrom the individual sites. This agrees with the strong variation between the sites visible in Figure 3, whereas a lower scatter can be seen in Figure 4. The differences between the sites emerge also in the values of the fitted parameters for the VWC dependence in the multi regression model. Since soil moisture is possibly related to soil texture and to the chemical characteristic of the soil, this outcome points to the relevance of different soil types. Different geological parental materials may result in different mineralogical composition of the soil, that in turn induces differences in bulk density (Brogowski et al., 2014; Mile and Mitkova, 2012), and consequently in soil porosity. Thus, water storage and percolation in the soil may be affected by the underlying geology through this chain of dependences. Future works will explore in detail the differences between the three sites, considering soil chemical and physical analyses.

Two types of sensitivity analysis were performed with MM3. First, the dataset was split in three subsets, excluding one year at a time from the calibration dataset. While models built using 2017 and either 2018 or 2019 were able to reproduced the excluded year (respectively, 2019 or 2018), the model built on 2018 and 2019 was not able to reproduce the fluxes recorded in the driest and warmest 2017. This suggests that the predictive ability of the model requires small variations of the drivers compared to those in the calibration dataset. Indeed, the average annual air temperature was much larger in 2017 compared to the two other years and variations of the average soil moisture ranged up to 30% over site A (Table 2). Notably, site A had also the largest RMSE in the predicted versus measured ER of 2017 (Table C.6 of the Appendix), while site C showed the lowest RMSE, as well as the lowest variation in air temperature and soil moisture. In general, the prediction of GPP for 2017 shows lower RMSE, reflecting the lower variations of *rs* compared to *Ta* with respect to the other years. The last observation confirms that the most relevant parameters are those related to the solar irradiance and environmental temperature. A bias in these parameters affects all the dependences, amplifying the effect. By contrast, a bias in the other predictors affects only one of the factors in the equation, limiting its influence to the specific term. Obviously, the larger and the more variegated is the dataset used for calibration, the stronger will be the model predictive ability.

Taking into the limitations discussed above, we estimated the possible $CO_2$ flux changes as a result of variations in mean temperature and soil water availability. Air temperature and soil humidity were varied, considering increases or decreases that were well within the range of intra-annual variations in the dataset, namely +0.5°C or +1 °C for temperature and $\pm 10\%$ of humidity. For variations of soil moisture at unchanged temperature, site A behaved differently compared to sites B and C, owing to the different sign of the term describing VWC in Eq.(8) and (9). As a result, both ER and GPP increased (decreased) in drier (wetter) conditions in site A, while the opposite was observed for sites B and C. Since soil moisture is potentially related to soil composition for the same water input, this result indicates that future projections should account for soil differences also at small scale.

When the effects of temperature and humidity changes were combined, a prevalence of the temperature variation was observed. As expected, ER was enhanced by the temperature rise, both in drier and wetter conditions for sites A and B, while site C showed a decreasing ER for higher temperature in drier conditions. This tendency is explained by a larger value of the parameter $a_1$ in site C compared to A and B, confirming that the results depend not only on the variation of the drivers, but also on the values of the parameters controlling the flux response (Tables 5 and 6).

The combined effect of temperature and soil moisture makes the assessment of possible future states strongly dependent on the ER/GPP ratio. For instance, a concomitant and similar trend of carbon emission and uptake in arid conditions was observed by Ciais et al. (2005), who suggested that some ecosystems could become carbon sources in dry climates, owing to a larger reduction in GPP compared to ER.

The systematic empirical approach discussed here is a versatile tool to unravel site-specific flux drivers at the watershed scale. Studies and assessments at watershed scale are often needed by stakeholders; empirical models with few parameters may be useful for short-time decision making. In any case, the results obtained for a specific environment should not be automatically transferred to other biomes or other climates, and projections obtained with empirical models are presumably informative only in the near future. Empirical models are not expected to provide meaningful results when the environmental conditions or the driver values are very different from those for which the model was established and tested. In particular, the application of our approach to other environments may identify different predictors, depending on the specific features of the sites.

Process-based models should probably be preferred for long-range forecast and generalizations, such as the DNDC model (e.g. Levy et al. 2007), PaSim (e.g. Ma et al., 2015), CHTESSEL (e.g. Boussetta et al, 2013), JULES (e.g. Van den Hoof et al., 2013) and Century (e.g. Conley et al. 2000). In this sense, process-based and empirical approaches are complementary. Empirical models are useful in the pioneering exploration of processes that are still poorly understood, relying on the possibility to obtain information directly from the data without, or with just a few, assumptions. This also includes processes whose functional dependence on the drivers is still missing or controversial. On the other hand, also process-based models include several unresolved processes that need to be parametrized. Empirical studies may then help to derive sensible parametrizations form data analysis.

The large representativeness of the data obtained with MM3 highlights the usefulness of this model compared to other empirical approaches. The systematic statistical effort reported here allowed to sort a model that was able to identify the most relevant drivers of $CO_2$ fluxes in dry grasslands. Interestingly, the drivers identified by MM3 agrees with the outcomes of other studies carried out in different regions of the world, pointing to some generality in the $CO_2$ flux controlling processes active in dry grasslands.

# 5  Conclusions

We studied carbon emission and fixation measured during the summers of 2017, 2018 and 2019 in an high-altitude Alpine meadow in north-western Italy. The systematic statistical analysis of the data allowed to identify carbon exchange drivers and establish an empirical model for $CO_2$ fluxes in mountain grasslands.

We compared three empirical modelling approaches based on different representations of additional environmental drivers on top of the standard ones (temperature for ER and solar irradiance for GPP). The best results in term of largest explained variance and higher efficiency were obtained using a model where the additional drivers acted as relatively small perturbations of the parameters in the standard functional regressions. The optimal set of predictors included environmental temperature, solar irradiance, soil moisture, air pressure and day of the year, in keeping with common observations on fluxes in dry grasslands. Both inter- and intra-annual variations of the fluxes were reproduced by the model, with the day of the year mainly accounting for the seasonal trend. Sensitivity analysis indicated that temperature and light are the most important controlling factors of the modelled fluxes, pointing to the need for their accurate estimation. Out-of-sample predictions helped to determine the applicability bounds of the model.

The comparison of the fluxes at three sites, with soils developed respectively over carbonate rocks, gneiss rocks and glacial deposits, indicated statistically significant differences in the carbon fluxes. This suggests that the soil physical and geochemical features should be taken into account in future studies, as an additional source of variability of carbon fluxes.

Application of the approach discussed here to other datasets can help providing a general *quantitative* model for the budgeting of the carbon cycle in climatic hot-spots and support the assessment of the possible triggers of new ecosystem states.

# Acknowledgements

The authors acknowledge Bruno Bassano and Ramona Viterbi of GPNP for help and support during the measurement campaigns, Massimo Guidi for discussions, advice and help and Gianluca Persia for participation during his Master Thesis. Maurizio Catania, Virginia Boiani, Stefano Ferraris, Elisa Palazzi and Maddalena Pennisi contributed to the measurement campaigns and made field work a moment of pleasant scientific discussion. The interaction with Tim White, Ashlee Dere and colleagues

__

during two summer schools at GPNP in 2018 and 2019 was a source of inspiration. This work was funded by the Project of Interest NextData of the Italian Ministry for Education, University and Research (MIUR) and by the EU H2020 ECOPOTENTIAL Project, grant number 641762. The authors acknowledge the anonymous reviewers for their useful remarks, that enhanced the presentation of our results.

# Appendix

## A. Instrumental Setup

We used a custom portable accumulation chamber to measure the soil gas fluxes, where the analytical instrumentation is placed inside a portable case equipped with a shoulder harness and straps (see Figure A.1). The block diagram of the device is the following:
- A cylindrical transparent (polycarbonate) accumulation chamber with open base area of 363 cm$^2$ and a height of 31.5 cm. A stainless collar (diameter 21.5 cm, height 5 cm) is used as interface between soil and the chamber;
- A pneumatic system comprising a pump and pipes to transport the soil gas from the chamber to the gas detectors and back to the chamber and a barometer;
- a LI-COR LI-840 single path, dual wavelength, nondispersive infrared (NDIR) carbon dioxide analyser (West Systems Ltd., 2002);
- A battery pack;
- A palmtop computer, including a GPS and connected to the LI-COR, to the barometer and to the other external sensors (air thermometer, radiometer, soil temperature and soil conductometer) through the Bluetooth wireless standard.

The environmental variables have been measured as follows:
- Air moisture ($q$), Air temperature ($Ta$) and solar irradiance ($rs$) are measured by an LSI Lastem thermoigrometer and an LSI Lastem ISO 9060 Class 2 pyranometer, respectively, at about 1.5 m above the ground, at the centre of the plot;
- Soil temperature ($Ts$) and soil volumetric water content ($VWC$), expected to have significant spatial heterogeneity, were measured at the base of the collar, with a PT100 sensor and an AT Delta-T SM150 T Soil Moisture sensor respectively, at a depth of about 5 cm.

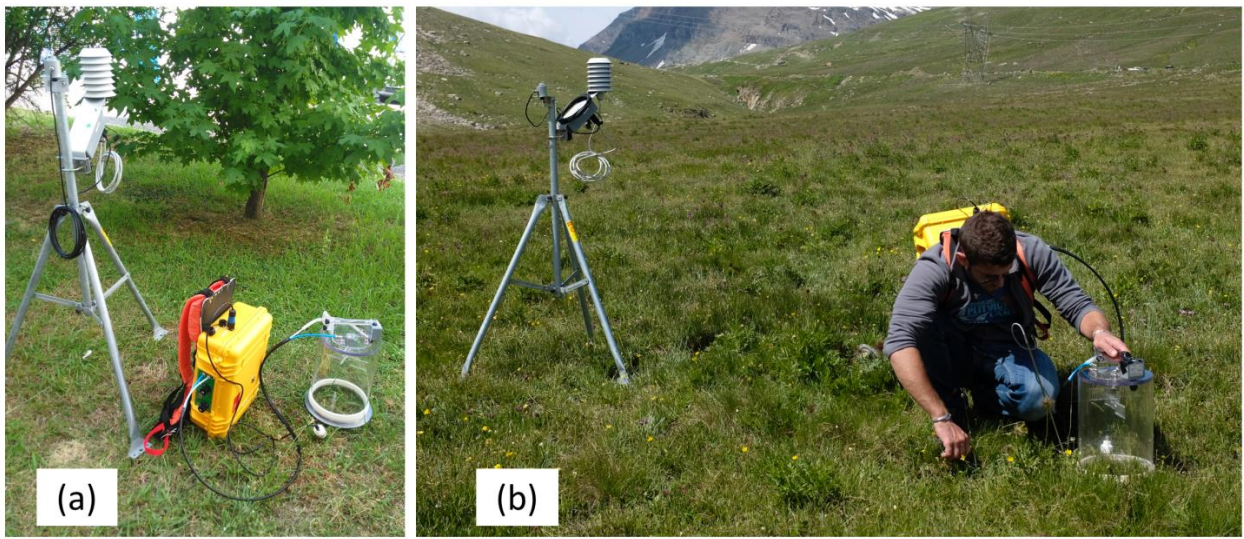

*Figure A.1: Instrumentation set up (a) and its use in the field (b).*

The flux $F$ of the gas of interest (e.g., $CO_2$) is obtained, as a first approximation, from the initial slope (at time zero) of the concentration-time curve ($dC/dt$), through the following equation (Chiodini et al., 1998):

$$F_{CO_2} = \frac{V}{A} \cdot \left(\frac{dC}{dt}\right)_{t=0} \cong H \cdot \left(\frac{dC}{dt}\right)_{t=0}, \tag{A.1}$$

where V [$m^3$], A [$m^2$] and H [$m$] are the volume, base area and height of the accumulation chamber, respectively. The value $(dC/dt)_{t=0}$ is measured in ppm/sec, and it is then converted to flux values ($mol/m^2/day$) by using the conversion factor ($mol \cdot sec/ppm/m^2/day$)

$$conversion\ factor = \frac{86400 \cdot P \cdot V}{10^6 \cdot R \cdot T \cdot A},$$

where:
R = 0.08314 *millibar·m³/K/mol*
P = barometric pressure expressed in *mbar*
T = air temperature expressed in Kelvin
V = chamber volume expressed in *m³*
A = chamber base area expressed in *m²*

Finally, a *day* to *second* and a *mol* to *μmol* conversions were applied during the data analysis, obtaining flux measurements in $\mu mol/s/m^2$. The dataset is available at
https://zenodo.org/record/3588380#.XgCwadZKhZ1 (Giamberini et al, 2019).

## B. Analytical Calculations

The dependence of the ER from the environmental temperature (*T*) has a standard exponential form of the type $ER = a\,e^{bT}$. In order to obtain a general formulation of MM3 let us assume that the parameters *a* and *b* depend on other variables. For shortness we will consider two variables named $x_1$ and $x_2$, different from *T*, although the following derivation can be extended to any set of additional variables. To a first instance, functions $a(x_1, x_2)$ and $b(x_1, x_2)$ can be expressed in a polynomial form as

$$a(x_2, x_3) \approx (a_0 + a_1 x_1 + a_2 x_2 + a_3 x_1^2 + a_4 x_2^2 + O(x_1^3) + O(x_2^3), \tag{B.1}$$

$$b(x_2, x_3) \approx (b_0 + b_1 x_1 + b_2 x_2 + b_3 x_1^2 + b_4 x_2^2 + O(x_1^3) + O(x_2^3). \tag{B.2}$$

Assuming $x_1$ and $x_2$ as small perturbations, Taylor expansion of the exponential is possible. Retaining only the first order of the expansion, i.e. $O(x_1)O(x_2)$, one obtains

$$ER = a(x_1, x_2)e^{b(x_1, x_2)T} \approx \tag{B.3}$$
$$\approx (a_0 + a_1 x_1 + a_2 x_2 + a_0 b_1 x_1 T + a_0 b_2 x_2 T)\,e^{b_0 T} + O(x_1^2) + O(x_2^2).$$

The same procedure was applied to the GPP dependence on light, namely $GPP = \frac{F_{max}\,\alpha\,rs}{F_{max} + \alpha\,rs}$. The parameters of the Michaelis-Menten function were expressed as follows

$$\alpha(y_1, y_2) \approx (\alpha_0 + \alpha_1\ y_1 + \alpha_2 y_2 + \alpha_3 y_1^2 + \alpha_4 y_2^2 + O(y_1^3) + O(y_2^3), \tag{B.4}$$

$$Fmax(y_1, y_2) \approx (F_0 + F_1\ y_1 + F_2 y_2 + F_3 y_1^2 + F_4 y_2^2 + O(y_1^3) + O(y_2^3). \tag{B.5}$$

Again, by expressing these functions in a polynomial form, using Taylor expansion one ends up with

$$GPP = \frac{F_{max}(y_1,y_2)\alpha(y_1,y_2)\,rs}{F_{max}(y_1,y_2) + \alpha(y_1,y_2)\,rs} \approx \frac{F_0\alpha_0 rs}{F_0+\alpha_0 rs}\left(1 + \frac{F_1\alpha_0 + F_0\alpha_1}{F_0\alpha_0}y_1 + \frac{F_2\alpha_0 + F_0\alpha_2}{F_0\alpha_0}y_2\right) + \\ -\left(\frac{F_0\alpha_0 rs}{F_0+\alpha_0 rs}\right)^2 \left(\frac{\alpha_1}{F_0\alpha_0}y_1 + \frac{\alpha_2}{F_0\alpha_0}y_2 + \frac{F_1}{F_0\alpha_0}\frac{y_1}{rs} + \frac{F_2}{F_0\alpha_0}\frac{y_2}{rs}\right) + O(y_1^2) + O(y_2^2)\ . \tag{B.6}$$

Then, Equation (4) is obtained by defining the parameters in Equation (5).

## C. Data Analysis

In this section we provide the supplementary information on the data analysis, with quantitative figures and tables supporting the results presented in the paper.

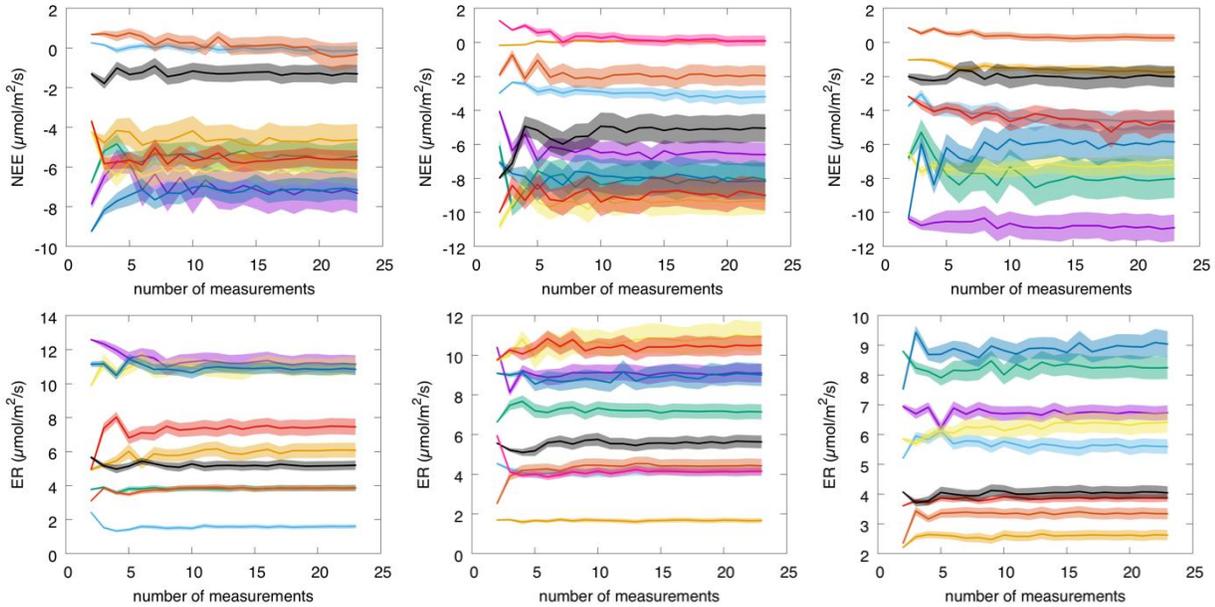

*Figure C.1. Means and standard deviations for each measurement campaign as a function of the number of individual sample point measurements included in the averaging procedure. Top panels are for NEE and bottom panels are for ER. Left panels are for site A (carbonate), center panels for site B (glacial) and right panels for site C (gneiss). Lines are weighted average over 23 groups of "k" elements (k varies in the range in [2,23]) that are randomly selected from all available measurements for that site and that measurement campaign (date). Shadows are the corresponding weighted standard deviations.*

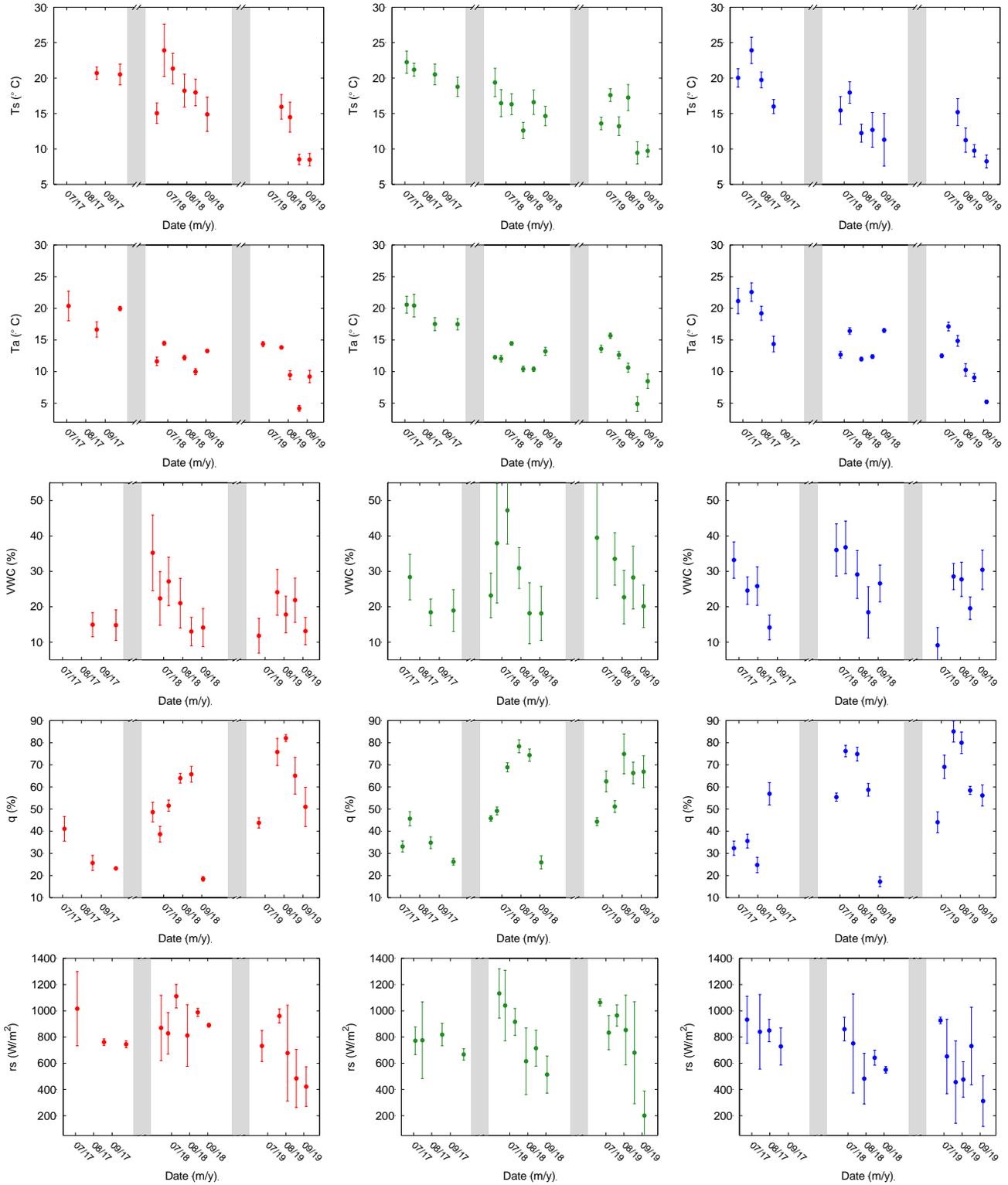

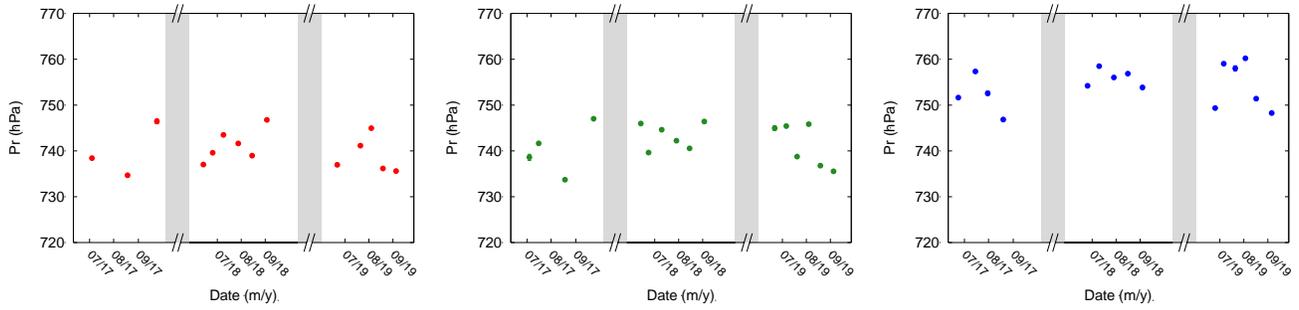

*Figure C.2. Values of the meteorological and environmental variables at site A (left, red), B (center, green) and C (right, blue) as a function of the measurement date. Error bars are 1 σ of the distribution of individual measurements. Ts = soil temperature at 5 cm; Ta = air temperature; VWC = volumetric water content; q = air moisture; rs = solar irradiance; Pr = atmospheric pressure.*

|   |           | Ts    | Ta    | VWC    | rs      | q     | Pr    | H     | DOY    | NEE   | ER    | GPP    |
|---|-----------|-------|-------|--------|---------|-------|-------|-------|--------|-------|-------|--------|
| A | 2019-2018 | **-6.11** | **-2.26** | **-4.41** | **-284.98** | **15.78** | **-2.27** | 0.05 | **6.80** | -0.05 | **-1.45** | 0.094 |
| A | 2019-2017 | **-8.82** | **-8.95** | 2.86 | **-205.35** | **33.03** | -0.09 | **0.55** | **-14.53** | -0.04 | 0.40 | -0.82 |
| A | 2018-2017 | **-2.58** | **-7.01** | **7.28** | **79.63** | **17.25** | **1.39** | **0.50** | **-21.33** | 0.008 | **1.85** | -1.77 |
| B | 2019-2018 | **-3.20** | **-1.23** | -0.008 | -54.15 | **4.95** | **-1.78** | -0.69 | 2.17 | 0.06 | **1.04** | -0.04 |
| B | 2019-2017 | **-7.88** | **-8.67** | **7.45** | 11.85 | **26.71** | **1.16** | -0.31 | **-11.08** | -0.05 | **2.64** | **-3.47** |
| B | 2018-2017 | **-4.67** | **-7.16** | **7.53** | 65.99 | **21.75** | **2.94** | 0.38 | **-13.25** | -1.11 | **1.92** | **-3.03** |
| C | 2019-2018 | **-2.65** | **-2.28** | **-6.26** | -55.98 | **9.43** | **-1.29** | 0.94 | -5.73 | 0.08 | 0.04 | -0.027 |
| C | 2019-2017 | **-8.64** | **-7.61** | -1.30 | **-237.78** | **28.53** | **2.49** | **1.37** | **6.16** | 24.18 | -0.01 | **2.52** |
| C | 2018-2017 | **-5.99** | **-5.33** | **4.95** | **-181.81** | **19.09** | **3.79** | 0.43 | **11.90** | 23.38 | -0.05 | **2.79** |

*Table C.1. Differences between years, with significant differences in bold (P<0.05). Air temperature(Ta) and soil temperature (Ts) are in °C, volumetric water content (VWC) and air moisture (q) in %, solar irradiance (rs) in W/m$^2$, air pressure (Pr) in hPa and fluxes in μmol/s /m$^2$ (GPP, NEE and ER are Gross Primary Production, Net Ecosystem Exchange and Ecosystem respiration, respectively). The hour of sampling (H) and day of the year (DOY), corresponding to the sampling date, vary in the interval 0-24h and 1-365d, respectively.*

|      |     | Ts     | Ta     | VWC     | rs       | q       | Pr       | H       | DOY    | NEE    | ER      | GPP     |
|------|-----|--------|--------|---------|----------|---------|----------|---------|--------|--------|---------|---------|
| 2017 | A-B | -0.06 | -0.003 | **-6.86** | **82.35** | **-4.95** | -0.40 | **-1.35** | 8.08 | 0.16 | 0.03 | 0.13 |
| 2017 | A-C | 0.69 | -0.32 | **-9.56** | 3.04 | **-7.39** | **-12.24** | **-1.05** | **24.83** | **2.01** | -0.26 | **2.27** |
| 2017 | B-C | 0.76 | -0.32 | **-2.70** | **-79.31** | -2.44 | **-11.83** | 0.30 | **16.75** | **1.85** | -0.29 | **2.14** |
| 2018 | A-B | **2.03** | 0.15 | **-7.11** | **95.99** | **-9.46** | **-1.95** | **-1.24** | 0 | **1.36** | -0.04 | 1.40 |
| 2018 | A-C | **4.11** | **-1.68** | **-7.24** | **265.49** | **-9.24** | **-14.64** | **-0.98** | **-8.4** | -0.24 | **2.05** | **-2.29** |
| 2018 | B-C | **2.07** | **-1.83** | -0.13 | **168.50** | 0.21 | **-12.69** | 0.25 | **-8.4** | **-1.60** | **2.09** | **-3.69** |
| 2019 | A-B | -0.87 | -0.88 | **-11.44** | **-134.84** | 1.37 | **-2.44** | **-0.49** | 4.63 | 0.25 | **-2.53** | **2.78** |
| 2019 | A-C | 0.65 | **-1.66** | **-5.39** | 35.48 | -2.90 | **-15.62** | **-1.87** | 4.13 | -0.82 | 0.25 | -1.07 |
| 2019 | B-C | **1.15** | -0.78 | **6.05** | **170.33** | **-4.27** | **-13.17** | **-1.37** | -0.5 | -1.07 | **2.78** | **-3.85** |
| All  | A-B | -0.004 | -0.49 | **-8.35** | 16.58 | **-3.65** | **-1.66** | **-0.97** | 2.9 | 0.65 | **-0.88** | **1.52** |
| All  | A-C | **1.34** | **-1.51** | **-6.45** | **122.27** | **-5.95** | **-14.24** | **-1.33** | 4.1 | 0.15 | **0.86** | -0.72 |
| All  | B-C | **1.34** | **-1.01** | **1.89** | **105.70** | -2.30 | **-12.58** | -0.37 | 1.2 | -0.50 | **1.74** | **-2.24** |

*Table C.2. Differences between sites, with significant differences in bold (P<0.05). Air temperature (Ta) and soil temperature (Ts) are in °C, volumetric water content (VWC) and air moisture (q) in %, solar irradiance (rs) in W/m$^2$, air pressure (Pr) in hPa and fluxes in μmol/s /m$^2$ (GPP, NEE and ER are Gross Primary Production, Net Ecosystem Exchange and Ecosystem respiration, respectively). The hour of sampling (H) and day of the year (DOY) vary in the interval 0-24h and 1-365d, respectively.*

| Site A | rs    | q     | Ta    | Pr    | VWC   | Ts    | DOY   | H     |
|--------|-------|-------|-------|-------|-------|-------|-------|-------|
| Rs     | **1.00** | -0.06 | 0.45  | 0.33  | 0.20  | **0.63** | -0.34 | 0.14  |
| q      | -0.06 | **1.00** | **-0.63** | -0.12 | 0.25  | -0.38 | -0.36 | -0.31 |
| Ta     | 0.45  | **-0.63** | **1.00** | 0.38  | -0.14 | **0.77** | 0.15  | 0.15  |
| Pr     | 0.33  | -0.12 | 0.38  | **1.00** | -0.23 | 0.25  | 0.31  | -0.04 |
| VWC    | 0.20  | 0.25  | -0.14 | -0.23 | **1.00** | -0.02 | **-0.84** | 0.13  |
| Ts     | **0.63** | -0.38 | **0.77** | 0.25  | -0.02 | **1.00** | -0.21 | 0.45  |
| DOY    | -0.34 | -0.36 | 0.15  | 0.31  | **-0.84** | -0.21 | **1.00** | -0.32 |
| h      | 0.14  | -0.31 | 0.15  | -0.04 | 0.13  | 0.45  | -0.32 | **1.00** |

| Site B | rs    | q     | Ta    | Pr    | VWC   | Ts    | DOY   | H     |
|--------|-------|-------|-------|-------|-------|-------|-------|-------|
| rs     | **1.00** | -0.14 | 0.25  | 0.28  | **0.52** | 0.45  | **-0.81** | -0.20 |
| q      | -0.14 | **1.00** | **-0.62** | -0.19 | 0.24  | -0.43 | -0.13 | -0.13 |
| Ta     | 0.25  | **-0.62** | **1.00** | 0.24  | 0.01  | **0.81** | -0.08 | -0.45 |
| Pr     | 0.28  | -0.19 | 0.24  | **1.00** | 0.11  | 0.29  | -0.14 | -0.32 |
| VWC    | **0.52** | 0.24  | 0.01  | 0.11  | **1.00** | -0.18 | **-0.68** | **-0.53** |
| Ts     | 0.45  | -0.43 | **0.81** | 0.29  | -0.18 | **1.00** | -0.18 | -0.16 |
| DOY    | **-0.81** | -0.13 | -0.08 | -0.14 | **-0.68** | -0.18 | **1.00** | 0.27  |
| h      | -0.20 | -0.13 | -0.45 | -0.32 | **-0.53** | -0.16 | 0.27  | **1.00** |

| Site C | rs    | q     | Ta    | Pr    | VWC   | Ts    | DOY   | H     |
|--------|-------|-------|-------|-------|-------|-------|-------|-------|
| rs     | **1.00** | -0.47 | **0.69** | -0.10 | 0.03  | **0.75** | **-0.63** | **-0.62** |
| q      | -0.47 | **1.00** | -0.53 | 0.41  | 0.11  | -0.34 | -0.10 | 0.25  |
| Ta     | **0.69** | -0.53 | **1.00** | 0.19  | 0.10  | **0.91** | **-0.55** | -0.30 |
| Pr     | -0.10 | 0.41  | 0.19  | **1.00** | 0.33  | 0.16  | -0.35 | -0.14 |
| VWC    | 0.03  | 0.11  | 0.10  | 0.33  | **1.00** | 0.14  | **-0.56** | -0.27 |
| Ts     | **0.75** | -0.34 | 0.91  | 0.16  | 0.14  | **1.00** | **-0.70** | -0.39 |
| DOY    | **-0.63** | -0.10 | **-0.55** | -0.35 | **-0.56** | **-0.70** | **1.00** | **0.62** |
| h      | **-0.62** | 0.25  | -0.30 | -0.14 | -0.27 | -0.39 | **0.62** | **1.00** |

*Table C.3. Correlation coefficients of meteo-climatic variables, computed on the mean values. Significant values, here marked in bold, were obtained by the shuffling method, using a double tailed test, assuming a significance level of 95%. The method tests the probability to obtain a larger correlation by chance, using the shuffling of the pairs $(x_i, x_j)$, with i and j randomly varying over the whole set of $x_i$ and $x_j$ values. The diagonal values are highlighted, although the shuffled pairs are never able to give correlation coefficient larger than 1, that is the unshuffled value, producing a P-value of 0 by definition. Air temperature (Ta) and soil temperature (Ts) are in °C, volumetric water content (VWC) and air moisture (q) in %, solar irradiance (rs) in $W/m^2$ and air pressure (Pr) in hPa. The hour of sampling (H) and day of the year (DOY) vary in the interval 0-24h and 1-365d, respectively.*

|   |   | MM1 | MM2 | MM3 |
|---|---|---|---|---|
| ER | A | -1.40 (0.32) | -3.43 (0.49) | -10.59 (0.85) |
|    | B | -2.89 (0.35) | -2.72 (0.42) | -25.86 (0.94) |
|    | C | -2.00 (0.33) | -4.70 (0.51) | -13.41 (0.83) |
| GPP | A | 0.34 (0.23) | 0.60 (0.32) | -13.92 (0.84) |
|     | B | -7.09 (0.50) | -6.78 (0.55) | -19.45 (0.88) |
|     | C | -7.33 (0.53) | -3.81 (0.48) | -20.31 (0.89) |

*Table C.4. AIC and explained variance (in brackets) of multivariate regressions using the three models.*

|  | parameter | Sites A-B | Sites A-C | Sites B-C |
|---|---|---|---|---|
| ER | $b$ [$°C^{-1}$] | 0.14 | 0.58 | 0.13 |
|    | $a_1$ [$\mu mol/m^2/s$] | **0.02** | 0.62 | 0.33 |
|    | $a_2$ [$\mu mol/m^2/hPa/s$] | 0.62 | 0.66 | 0.61 |
|    | $a_3$ [$\mu mol/m^2/s/d$] | 0.33 | 0.43 | 0.06 |
| GPP | $\alpha$ [$\mu mol/W/s$] | 0.97 | 0.97 | 0.99 |
|     | $A_1$ | **0.001** | **0.002** | 0.32 |
|     | $A_2$ [$1/d$] | 0.056 | 0.25 | 0.57 |

*Table C.5. Significance of the differences between the different sites of the model parameter values reported in Table 4. Significant values are highlighted in bold.*

| Projected Year | A | B | C |
|---|---|---|---|
| 2017 | 5.52 | 2.23 | 4.04 |
| 2018 | 2.96 | 1.79 | 1.21 |
| 2019 | 3.21 | 1.20 | 1.52 |

*Table C.6. RMSE of the out-of-samples projections for ER ($\mu mol/s/m^2$).*

| Projected Year | A | B | C |
|---|---|---|---|
| 2017 | 1.05 | 3.35 | 2.30 |
| 2018 | 4.37 | 3.24 | 1.87 |
| 2019 | 1.79 | 2.65 | 2.16 |

*Table C.7. RMSE of the out-of-samples projections for GPP ($\mu mol/s/m^2$).*